\documentclass[useAMS,usenatbib,psfig,eps,graphics]{mn2e}
\usepackage{natbib,psfig,graphicx, epsfig}

\def\grtsim{\mathrel{\hbox{\rlap{\hbox{\lower2pt\hbox{$\sim$}}}\raise2pt\hbox{$>$}}}} 
\def\lesssim{\mathrel{\hbox{\rlap{\hbox{\lower2pt\hbox{$\sim$}}}\raise2pt\hbox{$<$}}}}

\def\degree{\nobreak\ifmmode{^\circ}\else{$^\circ$}\fi}

\newcommand{\whzsr}{W~Hz$^{-1}$~sr$^{-1}$}

\def\msol{M$_{\odot}$}

\def\lmean{$\rho_{L_{\rm bol}}$}
\def\qmean{$\rho_{Q_{\rm jet}}$}
\def\qlmean{$\rho_{Q_{\rm jet}} /\rho_{L_{\rm bol}}$}
\def\amean{$ \hat{a}_{\rm fid} (z)  $}
\def\sp{$\hat{a}$}

\newcommand{\mbh}{$m_{\bullet}$}

\newcommand{\aap}{A\&A}

\newcommand{\apj}{ApJ}
\newcommand{\apjl}{ApJL}

\newcommand{\mnras}{MNRAS}

\newcommand{\prd}{Phys. Rev. D}

 \newcommand{\myemail}{alejo.martinez-sansigre@port.ac.uk}

\begin{document} 
 \topmargin -0.5in 

\title[Cosmic evolution in the spin  of SMBHs]{Evidence for cosmic evolution in the spin  of the most massive black holes}

\author[A. Mart\'\i nez-Sansigre \& S. Rawlings]{Alejo Mart\'\i nez-Sansigre$^{1,2,3}$\thanks{\myemail}, Steve Rawlings$^{2}$\\
$^{1}$Institute of Cosmology and Gravitation, University of Portsmouth, Dennis Sciama Building, Burnaby Road, Portsmouth, \\ PO1 3FX, United Kingdom \\
$^{2}$Astrophysics, Department of Physics, University of Oxford, Keble Road, Oxford OX1 3RH, United Kingdom \\
$^{3}$SEP{\it net}, South-East Physics network \\
 }
 
\date{}

\pagerange{\pageref{firstpage}--\pageref{lastpage}} \pubyear{}

\maketitle

\label{firstpage}

\begin{abstract}  
We use results from simulations of the production of
magnetohydrodynamic jets around black holes to derive the cosmic spin
history of the most massive black holes. We assume that the efficiency
of jet production is a monotonic function of spin \sp, as given by the
simulations, and that the accretion flow geometry is similarly thick
for quasars accreting close to the Eddington ratio and for
low-excitation radio galaxies accreting at very small Eddington rates.
We use the ratio of the comoving densities of the jet power and the
radiated accretion power associated with supermassive black holes with
\mbh$\grtsim$$10^{8}$~\msol to estimate the cosmic history of the
characteristic spin \sp.  The evolution of this ratio, which
increases with decreasing $z$, is consistent with a picture where the
$z$$\sim$0 active galactic nuclei have typically higher spins than
those at $z$$\sim$2 (with typical values $\hat{a}$$\sim$0.35-0.95 and
$\hat{a}$$\sim$0.0-0.25 respectively).  We discuss the implications in
terms of the relative importance of accretion and mergers in the
growth of supermassive black holes with \mbh$\grtsim$$10^{8}$~\msol.
 \end{abstract} 
 

\section{Introduction} \label{sec:intro}

Astrophysical black holes are described by two parameters, mass \mbh\,
and spin \sp, defining the structure of space-time within regions
close to the event horizon. Every massive galaxy has a supermassive black hole (SMBH) at its
centre, and the combination of accretion and coalescence of SMBHs of
similar mass, major merging, can lead to a wide spread in spins
\citep[e.g.][]{2008ApJ...684..822B}. 
A possible
constraint on the cosmic spin history is the ratio of  kinetic to radiated outputs in active
galactic nuclei (AGNs), since these are powered by SMBHs.

During accretion onto black holes, the amount of energy available for
radiation and for the production of jets is determined by the mass and
spin of the black hole, as well as the rate of accretion of matter and
the geometry of the accretion flow.  More massive black holes can
accrete more matter, but those that are spinning rapidly can extract
energy more efficiently from the infalling material
\citep[e.g.][]{1973blho.conf..343N,1977MNRAS.179..433B}. In addition,
geometrically-thick accretion flows can power jets more effectively
than thin ones because they are able to sustain more powerful poloidal
magnetic fields
\citep[e.g.][]{2001ApJ...548L...9M}.

In galactic black holes (GBHs), the presence of steady jets occurs
typically during low accretion rates (where by low we mean an
Eddington ratio $\lambda$$\equiv$$\dot{m}/\dot{m}_{\rm
  Edd}$$\lesssim$$10^{-2}$).  During transitions to moderate accretion
rates, $\lambda$$\grtsim$$10^{-2}$ the jet is temporarily enhanced in
power, and can then become steady, transient or totally absent
\citep{2004MNRAS.355.1105F}. This is interpreted as a transition from
advection-dominated accretion flows
\citep[ADAFs][]{1995ApJ...452..710N} that are geometrically thick and
can sustain significant jets, to gas-pressure-dominated accretion
flows which are thin, with aspect ratios (the ratio of height to
radius) $\sim$0.01 \citep{1973blho.conf..343N}.

Quasars and high-excitation radio galaxies are SMBHs accreting at a
high fraction of their Eddington limiting luminosity, so that the
accretion flow is geometrically thick
\citep[][note GBHs with $\lambda$$\sim$1 also produce continous jets, e.g. GRS 1915+105, Fender et al. 2004]{1973blho.conf..343N}.  Low-excitation radio
galaxies are SMBHs with very low accretion rates, ADAFs, that have
similarly thick accretion flows.  The
geometry is thus similar at both high and low accretion rates.

SMBHs are known to display a wide range of
jet powers with no observable differences in the accretion flow,
suggesting that a hidden variable must be controlling the jet
power. The `spin paradigm' assumes the black hole spin to be the
physical parameter controlling the kinetic output in the form of jets
\citep[e.g.][]{1977MNRAS.179..433B,2007ApJ...658..815S}.

GBHs show a similar wide spread in jet powers, but these are not found
to correlate with the published estimates of the spins
\citep{2010MNRAS.406.1425F}. If the spin estimates are correct, this
provides very stringent evidence against the spin paradigm for GBHs. However, due to the uncertainties in the measurements of both the black hole spin and the total jet power, the results found by    \citet{2010MNRAS.406.1425F} do       not provide robust evidence against the spin paradigm. The reader is referred to Section~7.2 of \citet[][hereafter MSR11]{2011arXiv1102.2228M} for a detailed discussion.

In this letter we make the assumption that spin is indeed an important
factor in powering jets, following the results of general-relativistic
magnetohydrodynamic simulations
\citep[e.g.][]{2006ApJ...641..103H}. Hence, assuming the spin paradigm
to be correct, we present an observational constraint on the evolution
of the mean spin of the most massive SMBHs as inferred from a measured
cosmic change in the ratio of the power output in jets to the power
radiated by active galactic nuclei (AGNs). We adopt a $\Lambda$CDM
cosmology with the following parameters: $h = H_{0} / (100 ~ \rm km ~
s^{-1} ~ Mpc^{-1}) = 0.7$; $\Omega_{\rm m} = 0.3$; $\Omega_{\Lambda} =
0.7$.

\section{Assumptions}

High accretion rates will lead to efficient radiation of the thermal
energy originating from viscous forces caused by magnetohydrodynamic
turbulence. A fraction $\epsilon$ of the accreted energy will be
radiated away, and the remaining energy will be advected into the
black hole. The bolometric power available for radiation can therefore
be described by:

\begin{equation}
L_{\rm bol} = \epsilon \dot{m}_{\bullet} c^{2} 
\end{equation}

 \noindent where $L_{\rm bol}$ is the bolometric luminosity due to
 radiation, $\dot{m}_{\bullet}$ is the rate of accretion of mass onto
 the SMBH, and the term $\epsilon$ is the radiative efficiency.

The shearing of magnetic fields frozen into the accreting plasma will
allow the extraction of electromagnetic energy to power jets. Close to
a rotating black hole, the dragging of inertial frames will lead to a
higher angular velocity (and hence larger shear) and additional
amplification of the magnetic fields occurs by extraction of energy
from the rotating black hole.  The power available for the production
of jets can be described as:

\begin{equation}
Q_{\rm jet} = \eta \dot{m}_{\bullet} c^{2}
\end{equation}

\noindent where $Q_{\rm jet}$ is the jet power and $\eta$ is the jet
efficiency \citep[e.g.][ this is also the same as $\epsilon_{\rm kin}$
  in Merloni \& Heinz 2008]{2006ApJ...641..103H}. Here it is assumed
that $\eta$ is a function of spin so that rapidly spinning black holes
will have a higher jet efficiency.

The AGNs concerned in this letter are powered by SMBHs accreting
either at very low or at very high fractions of their Eddington
limiting luminosity \citep[so either $\lambda$$\lesssim$$10^{-2}$ or
  $\lambda$$\sim$1][]{2001MNRAS.322..536W,2004MNRAS.351..347M,2009ApJ...696...24S}.
In these two limits of low- and high-accretion rates the geometry of
the accretion flows will be similar, with aspect ratios
$\grtsim$0.1. Hence, we assume that $\eta$ is a monotonic function of
\sp\, and that variations due to the accretion flow thickness are a
secondary effect (see also Section~2.2 of MSR11 for more details).

\section{The comoving jet power and radiated energy}

 To infer the cosmic spin history of SMBHs, we will use the mean
 comoving kinetic jet power density from AGNs producing jets, \qmean,
 and the mean comoving jet power from radiating AGNs, \lmean. These
 quantities can both be measured using:

\begin{eqnarray}
\rho_{Q_{\rm jet}} = \int Q_{\rm jet} \phi_{Q_{\rm jet}}(Q_{\rm jet},z) {\rm d}Q_{\rm jet} \\
\rho_{L_{\rm bol}}  = \int L_{\rm bol} \phi_{L_{\rm bol}}(L_{\rm bol},z) {\rm d}L_{\rm bol},
\end{eqnarray}

\noindent where $\phi_{Q_{\rm jet}}(Q_{\rm jet},z)={\rm d}^{2}N/{\rm
  d}Q_{\rm jet}{\rm d}z$ and $\phi_{L_{\rm bol}}(L_{\rm bol},z)={\rm
  d}^{2}N/{\rm d}L_{\rm bol}{\rm d}z$ are the luminosity functions for
$Q_{\rm jet}$ and $L_{\rm bol}$ respectively, representing the
comoving number density of AGNs with a given kinetic jet power or
bolometric luminosity, and their evolution with cosmological redshift
$z$

The jet power can be estimated from the low-frequency radio luminosity
density due to synchrotron radiation
\citep[e.g.][]{1999MNRAS.309.1017W}, assuming that the jet output
results in energy stored in radio source lobes together with
associated and ineluctable work done on the source
environment. Adopting standard assumptions for the way this energy is
shared between magnetic fields and particles, the work done inflating
these lobes as well as typical jet advance speeds and a typical
density profile for the intergalactic medium, \qmean\, can be
estimated from the radio luminosity function $\phi_{L_{\nu}}(L_{\nu
},z)$:
   
 \begin{eqnarray} 
 \left({ \rho_{Q_{\rm jet}}   \over {\rm W~Mpc^{-3}}} \right) =3\times10^{38} f^{3\over2}  
  \int\left( {L_{\nu} \over   10^{28}}\right)^{6\over 7} \phi_{L_{\nu}}(L_{\nu },z) {\rm d}L_{\nu}, 
 \label{eq:radio}
 \end{eqnarray}
 
\noindent where $L_{\nu}$ is the luminosity density at 151~MHz in
\whzsr.  The term $f$ represents the combination of several
uncertainty terms when estimating $Q_{\rm jet}$ from $L_{\nu}$ (for
example plasma volume filling factor or the fraction of energy in
non-radiating particles). An alternative method to estimating jet
powers is to estimate the energy stored in cavities in the X-ray
emitting intra-cluster gas, and both methods agree best for
$f$$\sim$20 \citep{2010ApJ...720.1066C}. We therefore initially take
$f$$=$20 but in Section~\ref{sec:cosmicspin} consider deviations from this assumption.

The radio luminosity function includes two components: the luminosity
function for the high radio luminosity classical double sources
\citep{2001MNRAS.322..536W} and the luminosity function for the lower
radio luminosity AGNs \citep{2009ApJ...696...24S}\footnote{The latter
  has been converted from 1.4~GHz to 151~MHz assuming
  $L_{\nu}$$\propto$$\nu^{-\alpha}$ with a radio spectral index of
  $\alpha$$=$0.75}.  For the lower radio luminosity AGNs, the
luminosity function extends to $z$$=$1.4 only, and we have
extrapolated it to $z$$=$2.0 using the same functional form for the
evolution. The luminosity functions of other AGNs suggest a turnover
around $z$$=$2, so no further extrapolation is attempted.  We neglect 
any errors in the quantity inside the integral in Equation~\ref{eq:radio} which is reasonable 
because it is essentially the luminosity density, a quantity dominated by objects near the break in the 
radio luminosity function: the normalization at this break is directly measured to $z \sim 2$ 
\citep{2001MNRAS.322..536W, 2010MNRAS.401.1709}.

 In AGNs the hard (high-energy) X-ray luminosity originates from
 inverse-Compton scattering of photons from the accretion flow by a
 corona of hot electrons and it is a good tracer of the bolometric
 luminosity.  The X-ray luminosity function $\phi_{L_{\rm X}}(\it
 L_{\rm X}, \it z)$ \citep{2008ApJ...679..118S} can be used to
 estimate \lmean:

\begin{equation} 
\left({ \rho_{L_{\rm bol}}   \over {\rm W~Mpc^{-3}}} \right)=  \int (1+ F_{\rm CT})\it  \,   C_{\rm X}  L_{\rm X} ~\phi_{L_{\rm X}} (\it L_{\rm X}, \it z){\rm d}L_{\rm X}. 
\label{eq:xray}
\end{equation}

\noindent The factor $F_{\rm CT}$ accounts for the fraction of
luminous Compton-thick AGNs missed by the hard X-ray surveys: 
\citep[$F_{\rm CT}$$\approx$0.5, e.g.
][]{2007MNRAS.379L...6M,2007A&A...463...79G},  and we adopt an uncertainty 
\citep[from Table 8 of][]{2009ApJ...693..447F}.
so that $F_{\rm CT} = 0.5 \pm  0.1$.
The bolometric
correction $C_{\rm X}$ converts from a monochromatic luminosity to a
bolometric luminosity, $L_{\rm bol}=C_{\rm X} L_{\rm X}$:  we use 
the values given by \citep{2007ApJ...654..731H} with, following
\citet{2009ApJ...692..964M}, an assumed 8 per cent uncertainty.

 The value of \qmean\, is always dominated
 by the output from the most massive SMBHs, with
 \mbh$\grtsim10^{8}$~\msol\, at all redshifts
 \citep{2004MNRAS.351..347M,2009ApJ...696...24S}.  To ensure that
 \lmean\, is also dominated by the most massive black holes, we only
 integrate above X-ray luminosities that can be achieved by SMBHs with
 \mbh$\geq$$10^{8}$~\msol\, accreting at $\geq$25\% of the Eddington
 accretion rate. This is a compromise between completeness to SMBHs
 with \mbh$\geq$$10^{8}$~\msol\, and contamination from lower-mass
 SMBHs. This value securely avoids serious contamination from SMBHs with
 masses $<$$10^{8}$~\msol. From the distribution in $\lambda$
 found by \citet[][their Figure 3]{2004ApJ...613..109H}  we estimate that it leads to
 \lmean\, at low-$z$ being underestimated by a factor $\sim$3. At high
redshift we expect it to underestimate \lmean\, by a negligible amount.

This evolution of the ratio \qlmean\, with redshift $z$ for the most
massive SMBHs (with \mbh$\grtsim10^{8}$~\msol) is shown in
Figure~1. There is likely some exaggeration of the rise in \qlmean\,
towards low $z$, because of the cut-off assumed in X-ray luminosity,
 but only by a factor $\sim$0.5 (base-10 logarithm), smaller than
  the inferred cosmic rise towards low $z$. Note also that the
absolute value of the height of the curve will scale with $f$ as
$f^{3/2}$.

\begin{figure} 
\begin{center}
\psfig{file=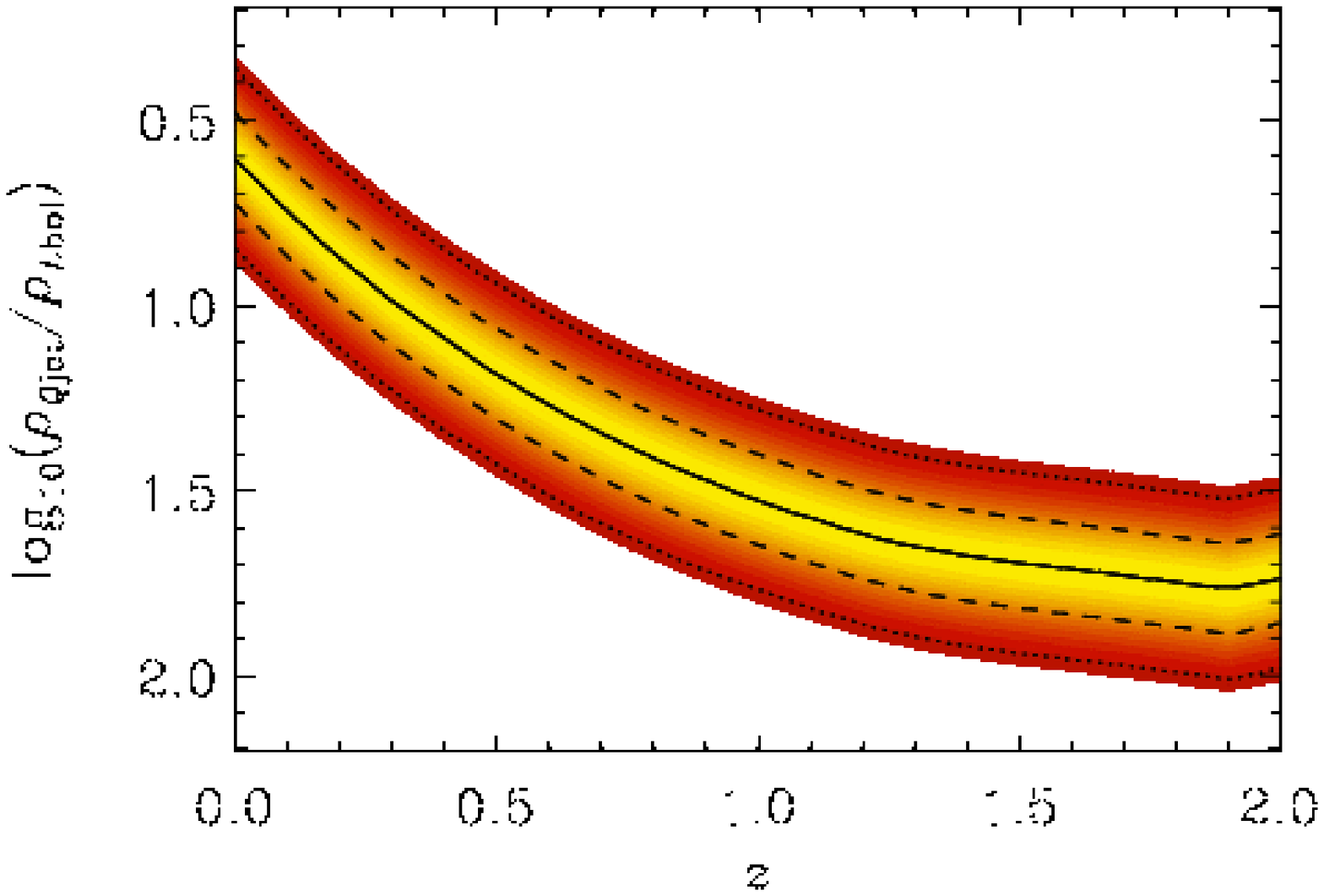, width=8cm, angle=0}
\caption{\noindent The posterior probability distribution function for
  the logarithm (base 10) of \qlmean\, as a function of redshift for
  the most massive black holes (with \mbh$\grtsim10^{8}$~\msol). The
  colours represent the normalised posterior probability of \qlmean,
  given the data. Light yellow represents the highest probabilities,
  dark red the lowest. Contours are also marked as black lines,
  including the line of maximum posterior probability (solid), as well
  as the $\pm$1$\sigma$ (dashed) and $\pm$2$\sigma$ (dotted) lines. At
  all redshifts we only integrate above X-ray luminosities that can be
  achieved by SMBHs with \mbh$\geq$$10^{8}$~\msol\, accreting at
  $\geq$25\% of the Eddington accretion rate \citep[the characteristic
    value found by][]{2004MNRAS.352.1390M}.  }
\label{fig:qlz} 
\end{center}
\end{figure}

\section{The cosmic spin history}
\label{sec:cosmicspin}

 The ratio of \qmean\, and \lmean, for the most massive black holes
 ($\grtsim10^{8}$~\msol) is then given by

\begin{equation} 
{\rho_{Q_{\rm jet}} \over \rho_{L_{\rm bol}} } = { \eta \rho_{\dot{m}_{\bullet Q_{\rm jet}}}    \over  \epsilon \rho_{\dot{m}_{\bullet L_{\rm bol}}}   }, 
\label{eq:ql}
\end{equation}

\noindent that includes the comoving density of accretion of the AGNs
contributing to $\rho_{Q_{\rm jet}}$, namely $\rho_{\dot{m}_{\bullet
    Q_{\rm jet}}}$ and of AGNs contributing to $\rho_{L_{\rm bol}}$,
namely $\rho_{\dot{m}_{\bullet L_{\rm bol}}}$.

Separating the contribution to \qmean\, from ADAFs and quasars,
$\rho_{Q_{\rm jet}}$$=$$\eta_{\rm ADAF} \rho_{\dot{m}_{\rm ADAF}} +
\eta_{\rm QSO} \rho_{\dot{m}_{\rm QSO}}$. In the case of \lmean,
although $\rho_{\dot{m}_{\rm ADAF}}$$\leq$$\rho_{\dot{m}_{\rm QSO}}$,
 $\epsilon_{\rm ADAF}$$\ll$$\epsilon_{\rm QSO}$, so that \lmean\, is
totally dominated by quasars and $\rho_{L_{\rm
    bol}}$$= $$\epsilon_{\rm QSO} \rho_{\dot{m}_{\rm QSO}}$.

Hence, 

\begin{equation}
{\rho_{Q_{\rm jet}}\over \rho_{L_{\rm bol}}} = \left(\eta_{\rm ADAF} \over  \epsilon_{\rm QSO} \right)\left(\rho_{\dot{m}_{\rm ADAF}} \over \rho_{\dot{m}_{\rm QSO}} \right) + \left(\eta_{\rm QSO} \over  \epsilon_{\rm QSO} \right).
\label{eq:two_terms}
\end{equation}

We consider two approaches to using \qlmean\, to constrain the cosmic
spin history. The first one is to attempt to use the ratio to find a
fiducial  spin, under some simplifying assumptions. The second is
to compare a parametric model to the observations, to see how
well it reproduces them.

In both approaches, the jet efficiency $\eta$ is described using the
results from a set of three-dimensional general-relativistic
magnetohydrodynamic simulations, which can be approximated as
\citep{2006ApJ...641..103H}:

\begin{equation}
\eta(|\hat{a}|) \approx 0.002 \left(1- | \hat{a} |\right)^{-1}.
\label{eq:hk06}
\end{equation}

\noindent For spin values \sp$=$ [0.0, 0.5, 0.998] the corresponding
jet efficiencies are $\eta$$=$[0.002, 0.004, 1.00].  However, our
results are very similar if we use other efficiencies from the
literature (see supplementary figures and MSR11).

For the radiative efficiency, we assume the
\citet{1973blho.conf..343N} model, which for spins \sp$=$ [0.0, 0.5,
  0.998] has values $\epsilon$$=$[0.057, 0.082, 0.321] for co-rotating
accretion\footnote{In Section~\ref{sec:dis}, we will discuss the
  relevance of chaotic accretion. In this paradigm, an average
  efficiency of co- and counter-rotating accretion should be used
  \citep{2008MNRAS.385.1621K}. The results of MSR11 used jet
  efficiencies from the literature for co-rotating accretion, so for
  consistency we limit ourselves to these.  The radiative efficiency
  is determined for quasars and their spins are close to \sp$=$0.0
  (MSR11) so the co- and counter-rotating radiative efficiencies are
  virtually identical. Therefore, although there is a slight
  conceptual inconsistency between our model radiative efficiency and
  our interpretation, in practice it makes no difference to the
  results.}.

\subsection{The fiducial cosmic spin}

At $z$$\grtsim$1, the density of accretion onto ADAFs is insignificant
compared to the density of accretion onto quasars \citep[e.g. see
  Figure~9 of MSR11 and Figure~4 of][]{2008MNRAS.388.1011M},
$\rho_{\dot{m}_{\rm ADAF}}$$\ll$$\rho_{\dot{m}_{\rm QSO}}$ so that
${\rho_{Q_{\rm jet}}/ \rho_{L_{\rm bol}}}$$=$$\eta_{\rm QSO} /
\epsilon_{\rm QSO}$. Therefore, at high redshift \qlmean\, gives a
good estimate of the ratio of efficiencies for quasars and, under the
spin paradigm, of their typical spin.

At very low redshifts the comoving density of accretion onto ADAFs is
comparable to that of quasars, so $\rho_{\dot{m}_{\rm ADAF}}$$\approx$
$\rho_{\dot{m}_{\rm QSO}}$ \citep[ MSR11]{2008MNRAS.388.1011M}, in
which case ${\rho_{Q_{\rm jet}}/ \rho_{L_{\rm bol}}}$$\approx$
$\left(\eta_{\rm ADAF}/\eta_{\rm QSO} + 1 \right)\eta_{\rm
  QSO}/\epsilon_{\rm QSO}$. It is therefore not straightforward to
interpret \qlmean\, in terms of typical efficiencies, but some
progress can be made under the assumption that $\eta_{\rm ADAF}
\grtsim \eta_{\rm QSO}$, so that \qlmean$\grtsim$$2\eta_{\rm QSO} /
\epsilon_{\rm QSO}$. This shows that at $z$$\sim$0 the ratio \qlmean\,
will overestimate the typical efficiencies of quasars and ADAFs.

Figure~2 represents a fiducial cosmic spin history of the most massive
black holes inferred from \qlmean. It represents the characteristic
spin if at a given redshift all SMBHs have the same spin. It is
derived by assuming the first term in Equation~\ref{eq:two_terms} is
negligible, so we expect this figure to provide an accurate estimate
at moderate and high redshifts, but to overpredict \amean\, at the low
redshifts, marked in gray.  Overestimating $\eta/\epsilon$ by a factor
of $>$2 at $z$$\sim$0 suggests that the typical spin at $z$$\sim$0 is
$<$0.9 rather than $\sim$0.95.  Consideration of $f$ suggests that
lower values of $f$ also lead to lower estimates of the typical
low-$z$ spin: re-running our analysis with $f=10$ (rather than $f=20$)
leads to no changes at high $z$ but a typical low-$z$ spin of 0.8. We
note that high values of $f$ are supported by X-ray cavity
observations undertaken at low radio powers and low redshift so lower
values of $f$ for the low-$z$ population are hard to envisage.  If
  we assume that different effects (e.g. $f$ and underestimating
  \lmean\, at low-$z$) add to an overall overestimate of \qlmean\, by
  a factor of $\sim$2 (4), this will result in a fiducial spin at
  $z$$\sim$0 of $ \hat{a}_{\rm fid}$$\sim$0.9 (0.7). In these cases
  the inferred evolution would be less extreme, but still present. 

\begin{figure} 
\begin{center}
\psfig{file=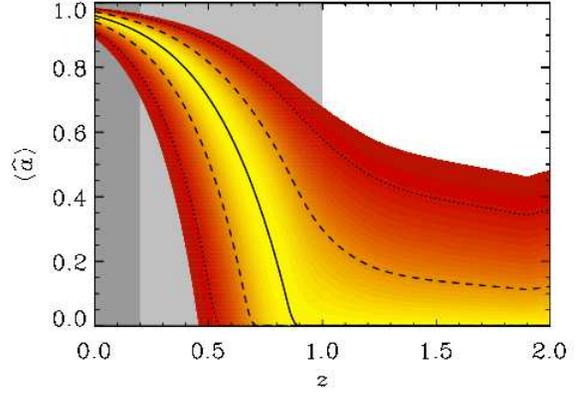, width=8cm, angle=0}
\caption{\noindent  Fiducial spin of the most massive black holes as a
  function of redshift. It shows the posterior probability
  distribution for \amean\, given the ratio \qlmean\, plotted in
  Figure~1, given $\eta$($|\hat{a}|$) and $\epsilon$($|\hat{a}|$).
  The posterior probability has been normalised at each redshift. The
  underlying assumption is that the aspect ratio of the accretion flow
  is similar for all the AGNs contributing to \qlmean, so that
  geometry has a negligible effect in the evolution of \qlmean, which
  is determined by the evolution in cosmic spin.  The colours and
  contours are the same as those in Figure~1.  At $z$$\grtsim$0.9, the
  maximum posterior probability line as well as the -1 and -2$\sigma$
  contours go along the \sp$=$0 axis. The grey shades mark the regions
  where the assumption that $\rho_{\dot{m}_{\rm ADAF}} \ll
  \rho_{\dot{m}_{\rm QSO}}$ breaks down.  In the light gray region, we
  estimate that $\rho_{\dot{m}_{\rm ADAF}}$$\approx$
  0.1$\rho_{\dot{m}_{\rm QSO}}$ while in the dark gray region,
  $\rho_{\dot{m}_{\rm ADAF}}$$\approx$ 0.25$\rho_{\dot{m}_{\rm QSO}}$
  (MSR11). Hence in these regions we overestimate \amean.  }
\label{fig:az} 
\end{center}
\end{figure}

To test wether this fiducial spin history is consistent with
constraints from the radiative efficiency of quasars
\citep[the][argument]{1982MNRAS.200..115S}, we compute the
luminosity-weighted spin (see also MSR11): $\langle
\hat{a}\rangle_{\rm L} \equiv \int  \hat{a}_{\rm fid}(z) 
\rho_{L_{\rm bol}} dz / \int \rho_{L_{\rm bol}} dz$.  The spin history
of Fig.~\ref{fig:az} yields a luminosity-weighted spin is $\langle
\hat{a}\rangle_{\rm L}$$=$0.055, which corresponds to a radiative
efficiency of $\langle \epsilon \rangle_{\rm L}$$=$0.059. This is in
good agreement with observational constraints, which typically yield
values $\langle \epsilon \rangle_{\rm L}$$\sim$0.05-0.10 \cite[see
  e.g.][]{2009ApJ...692..964M}.

\subsection{Can we explain \qlmean?}

In MSR11, we inferred a moderate evolution amongst the SMBHs with
\mbh$\geq$$10^{8}$~\msol. This evolution was really driven by the
change in relative space densities of two populations: the high
Eddington-rate SMBHs, or quasars, which dominate at high redshifts, were
found to have spin distributions centred around low
spin, \sp$\sim$0. The low Eddington-rate SMBHs, or ADAFs, become
significant at low redshifts, and these were found to have a bimodal
spin distribution, with a peak centred at \sp$\sim$0 and another at
\sp$\sim$1. These spin distributions allowed us to explain the local radio 
luminosity function, as well as explaining its evolution up to $z$$\sim$1. 

The results of MSR11 suggest that the evolution of the mean spin of
SMBHs is driven by a transition from a high-accretion rate, low spin
population, to a low-accretion rate population showing a bimodality in
spins (and hence a slightly higher mean spin). This evolution was
qualitatively similar to that seen in Fig.~\ref{fig:az}, but it was
significantly weaker, changing from $\langle \hat{a}
\rangle$$\sim$0.25 at high redshift to $\sim$0.35 at low redshift.

We can test whether the spin distributions we inferred in MSR11 can
reproduce the observed evolution in \qlmean. We use these spin
distributions to predict how the ratio \qlmean\, evolves with redshift
using

\begin{equation}
\left( { \rho_{Q_{\rm jet}}\over \rho_{L_{\rm bol}} }\right)_{\rm pred} = \left(\langle \eta_{\rm ADAF} \rangle \over  \langle \epsilon_{\rm QSO}\rangle \right)\left(\rho_{\dot{m}_{\rm ADAF}} \over \rho_{\dot{m}_{\rm QSO}} \right) + \left(\langle \eta_{\rm QSO}\rangle \over \langle \epsilon_{\rm QSO} \rangle \right).
\label{eq:pred_two_terms}
\end{equation}

\noindent Note that the angled brackets indicate expectation values:
$\langle \eta_{\rm ADAF} \rangle$ $=$ $\int \eta(\hat{a}) P_{\rm
  ADAF}(\hat{a})$d$\hat{a}$, $\langle \eta_{\rm QSO} \rangle$ $=$
$\int \eta(\hat{a}) P_{\rm QSO}(\hat{a})$d$\hat{a}$ and $\langle
\epsilon_{\rm QSO} \rangle$ $=$ $\int \epsilon(\hat{a}) P_{\rm
  QSO}(\hat{a})$d$\hat{a}$.  The variance of the ADAF term is large,
due to the bimodal nature, while the variance of the QSO term is
small.    As mentioned earlier, at $z$$\grtsim$1
$\rho_{\dot{m}_{\rm QSO}}$$\gg$$\rho_{\dot{m}_{\rm ADAF}}$ so that the
$\eta_{\rm QSO}$ term dominates, but at low redshift
$\rho_{\dot{m}_{\rm QSO}}$$\sim$$\rho_{\dot{m}_{\rm ADAF}}$.

\begin{figure} 
\begin{center}
\psfig{file=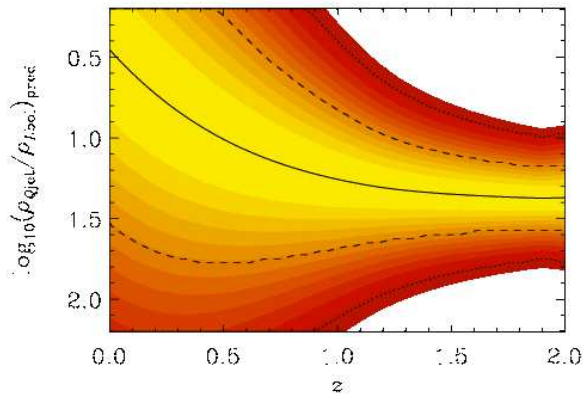, width=8cm, angle=0}
\caption{\noindent The evolution of \qlmean\, as predicted by the spin
  distributions from MSR11. In MSR11 we predicted spin distributions
  for high-Eddington rate objects (QSOs) and low-Eddington rate ones
  (ADAFs), as well as the accretion density onto these two
  sub-populations ($\rho_{\dot{m}_{\rm QSO}}$ and $\rho_{\dot{m}_{\rm
      ADAF}}$ respectively). Following Equations~\ref{eq:two_terms}
  and \ref{eq:pred_two_terms}, and the spin distributions from MSR11,
  we predict an expectation value that behaves very similarly to
  that of Fig.~\ref{fig:qlz}.  The contours are wide due to the
  variance of the spin probability distributions for high- and
  low-Eddington rate objects. At low redshifts the variance is larger
  due to the bimodality in the spin distribution of the low-Eddington
  rate objects.  }
\label{fig:pred} 
\end{center}
\end{figure}

Fig.~\ref{fig:pred} shows the resulting evolution of \qlmean, from
Equation~\ref{eq:pred_two_terms} and using the spin distributions from
MSR11. At $z$$\sim$0 the  most likely predicted value of \qlmean\, is
about 0.1 dex higher than the observed value, and at $z$$\grtsim$1 it
is about 0.3 dex higher. This can be understood since in MSR11
(Figs.~4 and 7 of that work) we overpredicted the radio luminosity
function of high-Eddington rate SMBHs, which dominate \qmean\, at high
redshifts.

The expectation value is systematically slightly higher
than the most likely value of \qlmean\, shown in Fig.~\ref{fig:qlz},
but the curvature is similar and the two figures agree very well
within their uncertainties.  Similar results are found using other jet
efficiencies from the literature (see supplementary figures).

Although the amount of evolution in the mean spin is different, Fig.~\ref{fig:az}
and the mean spin from MSR11, which can reproduce
Fig.~\ref{fig:qlz}, agree qualitatively: the typical spin was very
low at high redshifts, and has increased at low redshifts.

\section{Discussion}\label{sec:dis}

Using the observed ratio between the comoving densities of jet power
and radiation from SMBHs, \qlmean, we have inferred that the spin of
the SMBHs must be evolving. We have done this in two ways. The first
method included some simplifying assumptions to allow us to ``invert''
the problem. This allowed us to infer a fiducial spin for all SMBHs,
and we find this fiducial spin evolves strongly.  The second method
consists of comparing some highly parametrised description of the spin
distributions, inferred from fitting the local radio luminosity
function in our previous work (MSR11). There are no simplifying
assumptions, but rather than infer the evolution of a fiducial spin
value, we simply test whether  spin distributions  previously inferred in MSR11
can explain the observed evolution of \qlmean: we find that they can explain this evolution very  well.

The results from both methods are quantitatively different, in that
they differ in the  spin values inferred at low redshift. However the two methods agree
qualitatively: a low-spin epoch occurs at $z\grtsim1$ and coincides
with the period during which the black holes with masses
\mbh$\grtsim10^{8}$~\msol\, accreted most of their mass
\citep{2002MNRAS.335..965Y}. This epoch is one characterised by high
accretion rates onto low-spin objects, while the present-day epoch
($z$$\sim$0) is characterised by low accretion rates onto a population
of SMBHs where a significant fraction have high spins.

Somewhat counter-intuitively, the epoch of lowest-spin corresponds to
the epoch of highest accretion.
This, however,  is consistent instead with the  paradigm of ``chaotic accretion"  \citep[e.g.][]{2008MNRAS.385.1621K}: approximately half of the accretion occurs  
  with the disc counter-rotating with respect to the SMBH, hence decreasing the
  spin of the hole as opposed to the increase brought by a co-rotating disc.
  
 Chaotic accretion is expected to lead to SMBHs with low spins,
 $\hat{a}$$\sim$0.1 \citep{2008MNRAS.385.1621K}, and our result is
 thus in agreement with black holes being ``spun down" at high
 redshift, when the typical accretion rate was higher, due to the
 greater available supply of concentrated cold gas in the central
 regions of the host galaxy \citep[e.g.][]{2009ApJ...696L.129O}.

  When two black holes of similar mass coalesce, the orbital angular
  momentum contributes significantly to the final angular momentum of
  the coalesced black hole, leading to high values of the final spin,
  typically \sp$\sim$0.7
  \citep{2008PhRvD..78d4002R}.  Hence, mergers of
  SMBHs of similar mass (major mergers, with mass ratios $\lesssim$4:1)
  are expected to provide the ``spin-up" mechanism.

  SMBHs of mass $\grtsim$$10^{8}$~\msol\, are hosted by galaxies with 
  bulge masses $\grtsim$$10^{11}$~\msol, typically elliptical galaxies. 
  A significant fraction of these galaxies have undergone major mergers
  since $z\sim1$ \citep{2010ApJ...719..844R}, meaning their central SMBHs 
  will have most likely undergone a major merger too.

Due to the steepness of
the mass function above the break in the mass function,  our result is 
most likely dominated by the behaviour of SMBHs hosted by 
galaxies with masses $\sim$$1\times10^{11}$~\msol\, and reflects
their merger history, rather than 
reflecting that of the rarer, more massive galaxies with $>$$3\times10^{11}$~\msol, 
which seem to have undergone an earlier evolution \citep{2010MNRAS.402.2264B}.

  The mechanism for spinning-up SMBHs, major merging, is present
  during the entire epoch of $0 \leq z \leq 2$ since the merging of
  galaxies with similar masses is also a common occurrence at
  redshifts $z$$\sim$2. However, the centrally concentrated, typically
  molecular, cold gas content of the merging galaxies, the fuel for
  accretion, decreases significantly with cosmic time. 
  
  It is therefore the gradual disappearance of the
  ``braking'' mechanism, chaotic accretion, which determines the
 increase in the typical spin of the most massive black holes.

\smallskip

 We thank R.~Houghton, T.~Mauch, D.~Obreschkow, A.~Robaina and
 S.~Shabala and the anonymous referee for useful discussions.  A. M.-S. gratefully acknowledges
 an STFC Post-Doctoral Fellowship, reference ST/G004420/1. This work
 was also partly supported by the EC FP6, SKADS.

\bibliographystyle{bibstyle}

\label{lastpage}

 \end{document}